\documentclass[published]{JHEP3}
\JHEP{07(2003)033}

\skip\footins = 1\bigskipamount plus 2pt minus 4pt

\usepackage{epsfig}

\newcommand{\SU}{\mathop{\rm SU}\nolimits}
\newcommand{\UU}{\mathop{\rm {}U}\nolimits}
\newcommand{\Tr}{\mathop{\rm Tr}\nolimits}
\newcommand{\MeV}{\,{\rm MeV}}
\newcommand{\fm}{\,{\rm fm}}
\def\gsi{\gtrsim}
\newcommand{\eps}{\epsilon}
\newcommand{\la}{\langle}
\newcommand{\ra}{\rangle}

\title{Spectral properties of the overlap Dirac operator in QCD}

\author{Wolfgang Bietenholz and  Stanislav~Shcheredin\\
        Institut f\"{u}r Physik, Humboldt Universit\"{a}t zu Berlin\\
	Newtonstr.\ 15, D-12489 Berlin, Germany\\
	E-mail: \email{bietenho@physik.hu-berlin.de},
	        \email{shchered@physik.hu-berlin.de}}

\author{Karl Jansen\\
        NIC/DESY Zeuthen\\ 
	Platanenallee 6, D-15738 Zeuthen, Germany\\
	E-mail: \email{karl.jansen@desy.ch}}

\abstract{We discuss the eigenvalue distribution of the overlap Dirac
operator in quenched QCD on lattices of size $8^{4}$, $10^{4}$ and
$12^{4}$ at $\beta = 5.85$ and $\beta = 6$. We distinguish the
topological sectors and study the distributions of the leading
non-zero eigenvalues, which are stereographically mapped onto the
imaginary axis. Thus they can be compared to the predictions of random
matrix theory applied to the $\epsilon$-expansion of chiral perturbation
theory. We find a satisfactory agreement, if the physical volume
exceeds about $(1.2\fm)^{4}$.  For the unfolded level spacing
distribution we find an accurate agreement with the random matrix
conjecture on all volumes that we considered.}

\received{July 8, 2003}
\accepted{July 15, 2003}

\begin{document}

\section{Introduction}

\emph{Chiral perturbation theory} is a powerful tool to analyze QCD at
low energy. In this context, the pions as the lightest hadrons carry
the relevant degrees of freedom. They are considered as
quasi-Nambu-Goldstone bosons obtained from chiral symmetry breaking
--- their mass is provided by the masses of the $u$ and $d$ quark,
which are small compared to $\Lambda_{\rm QCD}$. One then constructs
an effective theory for the expansion in the pion mass and momenta,
which is strongly constrained by the requirements of chiral symmetry.

In particular the so-called \emph{$\eps$-expansion} is designed to
describe the system in a volume which is so small that it cannot even
include a pion Compton wave length~\cite{GasLeu}.  In this situation
the r\^{o}le of the topological sectors of the gauge field is far more
important than it is the case in a large volume~\cite{LeuSmi}.

In view of the comparison to lattice data, it is an important virtue
that a finite and even small volume can be interpreted directly --- in
contrast to most other lattice simulations one does not need to
extrapolate to large volumes.  Chiral perturbation theory in the
$\eps$-expansion describes the finite volume and quark mass dependence
of various quantities such as the scalar condensate or 2-point
correlation functions. These dependencies are parametrized by
\emph{the same} low energy constants as in the effective chiral
lagrangian in the infinite volume.  Therefore the unphysical small
volume can provide results for physical parameters.  The finite size
effects allow for a determination of the low energy constants in the
effective chiral lagrangian.  Examples of this procedure are given in
ref.~\cite{o4} for the O(4) symmetric non-linear $\sigma$-model, and
in ref.~\cite{condensate} for (quenched) QCD.
                
In this paper we are going to investigate \emph{Ginsparg-Wilson
fermions}~\cite{GW,Has}, which allow for simulations at small fermion
masses.  In particular, the evaluation of the corresponding lattice
Dirac operator can be done directly in the chiral limit, where the
Ginsparg-Wilson fermion has an exact (lattice modified) chiral
symmetry at finite lattice spacing \cite{ML}.  Moreover, since the
Ginsparg-Wilson fermions have exact zero modes, we can use them to
define the topological charge of the lattice gauge field~\cite{Has,ML}
by means of the Atiyah-Singer Index Theorem~\cite{AS}.

A particularly simple solution of the Ginsparg-Wilson relation is
obtained if the Wilson operator is inserted into the overlap
formula~\cite{HN}. As long as the gauge coupling is not too strong,
such a lattice Dirac operator is local~\cite{HJL} in the sense of an
exponential localization, hence it is conceptually correct in view of
the continuum limit.  However, its simulation is tedious: for the time
being, only quenched QCD simulations are feasible.  Since also the
quenched chiral perturbation theory in the $\eps$-regime has been
analyzed~\cite{DDHJ}, the simulation data can be compared to these
analytical predictions.

Even without considering hadrons, there are interesting predictions
for the spectrum of the Dirac operator which are based on Random
Matrix Theory (RMT), following the ideas of ref.~\cite{LeuSmi}. Chiral
RMT simplifies QCD to a gaussian distribution of the elements in the
fermion matrix, with the global symmetries of the QCD Dirac
operator. For a review on this field, see for instance
ref.~\cite{review}.  We only mention that although on the lagrangian 
level RMT is equivalent to chiral perturbation theory at lowest order
in the $\epsilon$-expansion, the predictions for the eigenvalue
distributions involve additional assumptions. Thus numerical
simulations can provide a test of the predictions from RMT beyond
chiral perturbation theory. Even if the fermion mass vanishes, there
is still a \emph{lower limit} for the volume where RMT applies; in
some steps, the latter assumes the volume (mathematically) to be
infinite. Theoretically one often requires the temperature $T$ to be
well below the Thouless energy $T \ll E_c = F_{\pi}^{2}/ ( \sqrt{V}
\Sigma )$.  However, it is not predicted explicitly at which lower
bound for the volume RMT really collapses.

There are explicit predictions for the statistical distribution of the
low-lying eigenvalues~\cite{lowEV,lowEV2} which can be confronted with
lattice data. We refer here to predictions in full QCD, although one
might be worried that quenching causes logarithmic
corrections~\cite{loga}. In this comparison, the chiral condensate
$\Sigma$ enters as a free parameter, hence, in turn, the fit to the
functions predicted by RMT provide a value for $\Sigma$.  Another
theoretical conjecture refers to the unfolded level spacing
distribution which is not sensitive to the energy scale, but which
allows for the inclusion of bulk eigenvalues.

Such studies were performed before, using staggered fermions in
QCD~\cite{stagger}, as well as Ginsparg-Wilson fermions in the
Schwinger model~\cite{Graz}, in QED~\cite{QED} and in QCD on $4^4$
lattices~\cite{SCRI,Bern}.  Here we are going to present data for the
overlap Dirac operator spectrum in quenched QCD on larger lattices
than used before for Ginsparg-Wilson fermions.

\section{The spectrum of the overlap Dirac operator
compared to the RMT predictions}

\subsection{Theoretical aspects}

In QCD with two massless quark flavors, the chiral symmetry
breaks spontaneously as
\begin{equation}
\SU(2)_{L} \otimes \SU(2)_{R} \to \SU(2)_{L+R} \,,
\end{equation}
which generates three Nambu-Goldstone bosons. If we deal with light
$u$ and $d$ quarks, these bosons pick up small masses because the
quark mass breaks the chiral symmetry explicitly.  The resulting
quasi-Nambu-Goldstone bosons are identified with the pions. They can
be represented by $\SU(2)$ matrix fields $U(x)$. At low energy
they are effectively described by chiral perturbation theory. To the
leading order, the corresponding lagrangian reads
\begin{equation}
{\cal L}[U] = \frac{F_{\pi}^{2}}{4}  \Tr 
\Big[ \partial_{\mu} U(x) \partial_{\mu} U^{\dagger}(x) \Big]
- \frac{\Sigma m}{2} \Tr  \Big[ e^{i\theta /2} U(x) +
 e^{-i\theta /2} U^{\dagger}(x) \Big] \, . 
\end{equation}
Here $m$ is the quark mass (for simplicity we assume it to coincide
for the two flavors) and $\theta$ is the vacuum angle.  The coupling
constants in leading order are the pion decay constant $F_{\pi}$ and
the chiral condensate $\Sigma$.

We put the system in a periodic box of size $L^{4}$.  Then the
$\eps$-regime is characterized by the condition
\begin{equation}
\frac{1}{m_{\pi}} \gg L \gg \frac{1}{4\pi F_{\pi}} \,, \qquad
m_{\pi}  =  \hbox{pion mass} \,.
\end{equation}

In this regime, RMT can be applied to the Dirac operator
$D$. In the continuum it leads to explicit formulae
for the statistical distribution of the low eigenvalues,
i.e.\ for the \emph{``microscopic regime''}. There the momenta
are counted as $O(1/L)$.
One introduces the spectral density
\begin{equation}
\rho(\lambda ) = \left\la \sum_{n} \delta (\lambda - \lambda_{n})\right\ra,
\end{equation}
where the sum runs over all modes, and $\lambda_{n}$ are the
eigenvalues of the operator $iD$.  It is favorable to use the
dimensionless variable
\begin{equation}
z = \lambda V \Sigma \,,
\end{equation}
which is of $O(1)$.  In these terms, the RMT prediction addresses the
microscopic spectral density
\begin{equation}
\rho_{s}(z) = \ ^{~ \lim}_{V \to \infty} \frac{1}{V \Sigma} \, \rho
\left(\frac{z}{V\Sigma}\right),
\end{equation}
which can be decomposed into the contributions of different
topological sectors,
\begin{equation}
\rho_{s} (z) = \sum_{\nu = -\infty}^{\infty} \rho_{s}^{(\nu )}(z) \,,
\qquad \nu = \hbox{topological charge}\,.
\end{equation}
Refs.~\cite{lowEV2} present expressions for the leading contributions
$\rho^{(\nu )}_{n}(z)$ in
\begin{equation}
\rho_{s}^{(\nu )}(z) = \sum_{n}{'} \rho^{(\nu )}_{n}(z) \,,
\end{equation}
where the sum runs over the non-zero modes.  The predictions for the
lowest eigenvalue at $\vert \nu \vert = 0,$ 1 and 2 are depicted in
figure~\ref{RMT}.  We see in particular that the density peak moves to
larger values of $z$ as $\vert \nu \vert$ increases.
\FIGURE[t]{\centerline{\epsfig{file=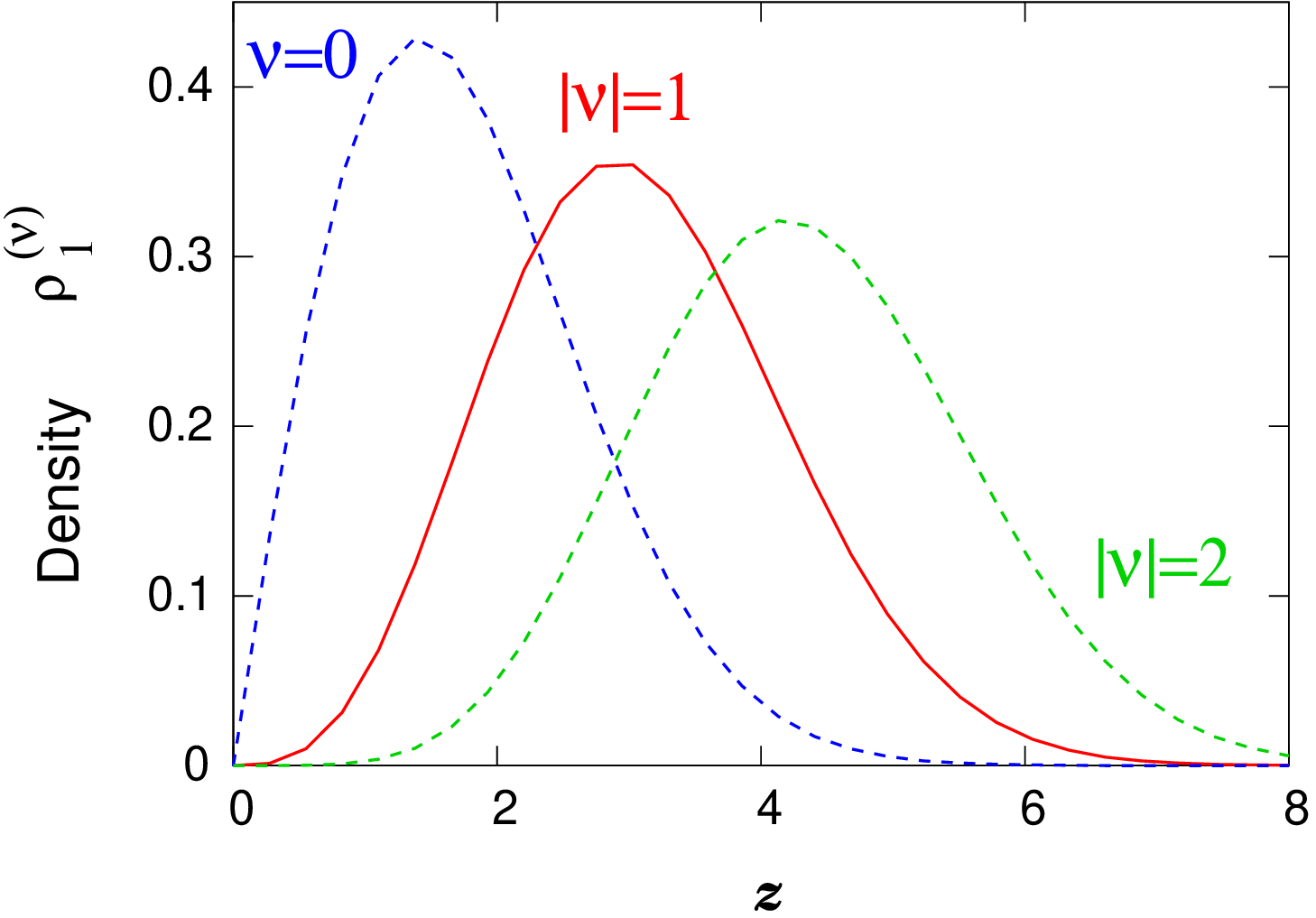,angle=270,width=.6\textwidth,clip=}}%
\caption{The distributions for the lowest eigenvalue at $\vert \nu
\vert = 0,$ $1$ and $2$, as predicted by chiral
RMT.\label{RMT}\label{fig1}}}

\subsection{Lattice results in the microscopic regime}

In order to verify the viability of the $\eps$-expansion and of chiral
RMT, it is of interest to compare these predictions to the
corresponding eigenvalue distributions obtained from lattice
simulations. The Wilson Dirac operator seems hardly suitable in this
respect; because of the mass renormalization it would be highly
problematic to identify the zero modes and the relative values of the
remaining modes.  In addition, simulations at small quark masses are
problematic due to exceptional configurations.
 
Staggered fermions do not suffer from additive mass renormalization
because they have an exact remnant chiral symmetry $\UU(1) \otimes
\UU(1)$.  Indeed such comparisons exist.  The staggered fermion
spectrum agrees well with the RMT prediction in the sector $\nu
=0$. However, it turns out that all other sectors, $\nu \neq 0$, yield
the same distributions, in particular the same histogram for
$\rho^{(\nu )}_{1}(z)$ in QCD~\cite{stagger}. It seems that staggered
fermions are generally insensitive to the topology, and therefore not
adequate for this purpose, at least for moderate and strong gauge
coupling.

Another type of lattice fermions without additive mass renormalization
is the Ginsparg-Wilson fermions.  Their lattice Dirac operator $D$
fulfills the Ginsparg-Wilson relation (GWR), which reads (in its
simplest form)
\begin{equation}  \label{GWR}
D \gamma_{5} + \gamma_{5} D = \frac{a}{\mu} D \gamma_{5} D \,,
\end{equation}
i.e.\ the continuum condition for the right-hand-side to vanish is
relaxed to a term of $O(a)$, where $a$ is the lattice spacing.  $\mu$
is a mass parameter, see below.  The GWR implies that $D^{-1}
\gamma_{5} + \gamma_{5} D^{-1} = a \gamma_{5}/\mu$ is local, therefore
the poles in the propagator are not shifted away from zero. Thus the
Ginsparg-Wilson fermions have exact zero modes, which occur with
positive or negative chirality.  We now define the topological charge
$\nu$ simply by the index
\begin{equation}
\nu = n_{+} - n_{-} \,,
\end{equation}
where $n_{+}$ ($n_{-}$) is the number of positive (negative) chiral
zero modes. \pagebreak[3] Here one adapts the continuum Index Theorem and uses it to
define the topological charge of the lattice gauge
configurations~\cite{Has,ML}.  The applicability of chiral RMT to
Ginsparg-Wilson fermions has been discussed theoretically in
ref.~\cite{split}.

Many lattice Dirac operators, such as the Wilson operator $D_{W}$,
obey $\gamma_{5}$ Hermiticity, $D_{W}^{\dagger} = \gamma_{5} D_{W}
\gamma_{5}$. This allows for a simple solution of the GWR by inserting
$D_{W}$ (say with Wilson parameter 1) into the overlap
formula~\cite{HN},
\begin{equation}
D_{\rm ov} = \frac{\mu}{a} \left[ 1 + {A}/{\sqrt{A^{\dagger}A}}\right],
\qquad A = aD_{W} - \mu \,.  \label{ovop}
\end{equation}
$D_{\rm ov}$ is a solution to condition~(\ref{GWR})
and it is $\gamma_5$-hermitean as well. (There are
alternatives to replace $D_W$ by lattice Dirac 
operators that are more involved but 
lead to improved properties, as suggested in refs.~\cite{HF1}.)
$\mu $ represents a negative mass of the Wilson fermion, which 
can be chosen in some interval as long as the gauge fields are smooth.
In the free case $\mu =1$ is optimal with respect to locality,
but at $\beta =6$ and $\beta = 5.85$ we move to $\mu = 1.4$
resp.\ $\mu = 1.6$ to compensate the mass renormalization of the 
Wilson fermion.

Note that the operator $A/\sqrt{A^{\dagger}A}$ is unitary, hence the
spectrum of $D_{\rm ov}$ is located on a circle in the complex plane
through zero, with center and radius $\mu /a$.  In order to relate the
eigenvalues found on this Ginsparg-Wilson circle to the continuum
eigenvalues $\lambda_{n}$, we map the circle stereographically onto
the imaginary axis. Requiring $f(z) = z + O(z^{2})$ and $f(2\mu ) =
\infty$ singles out the M\"{o}bius transform
\begin{equation}
f(z) = \frac{z}{1 - z/(2\mu )} \,.
\end{equation}
This mapping has been suggested before in ref.~\cite{stereo}, in
connection with the Leutwyler-Smilga sum rules~\cite{LeuSmi}.

The eigenvalues of Ginsparg-Wilson operators were compared to the RMT
prediction for the Schwinger model in ref.~\cite{Graz}, where the
overlap operator~(\ref{ovop}) as well as a truncated fixed point Dirac
operator were considered. In both cases the results agreed with the
RMT formulae.

Such studies were also performed with the overlap operator in 4d
QED~\cite{QED} and again the predictions were successful within the
statistical errors.

Finally QCD was considered, but only on small lattices of size
$4^{4}$. Ref.~\cite{SCRI} used the overlap operator at strong coupling
of $\beta = 5.1$.  Ref.~\cite{Bern} applied a truncated fixed point
action and obtained a decent agreement in a volume of $V= (1.2 \fm)^{4}$.  However, when the physical lattice spacing is decreased
so that the volume shrinks to $(0.88 \fm)^{4}$, the leading
non-zero eigenvalue distributions of the different topological sectors
are on top of each other, in contrast to the RMT prediction.

Our results are obtained by using again the overlap
operator~(\ref{ovop}) in QCD with the standard plaquette gauge action.
We have chosen a rather weak gauge coupling, $\beta \geq 5.85$, so
that the overlap formula can safely be applied to construct a
Ginsparg-Wilson fermion.\footnote{From our point of view, it is not
clear if the overlap fermion is well-controlled at stronger coupling.
For instance, typical spectra at $\beta = 5$ and even $\beta=5.4$ on
$4^4$ lattices have their eigenvalues spread over a broad
area~\cite{China}, so that it is hardly possible to find a good value
for $\mu$ which would split the (nearly) real eigenvalues into a small
branch (to be mapped to the vicinity of 0), and a large branch (to be
mapped onto the opposite arc of the Ginsparg-Wilson circle).}  Due to
the computational effort required by the overlap operator we had to
use the quenched approximation, as it was also the case in all
previous studies mentioned above.  We approximated the inverse square
root by Chebyshev polynomials to an accuracy of $10^{-12}$. Then we
used the PARPACK routines to evaluate up to 100 eigenvalues of
$D_{\rm ov}$.  We focused on the eigenvalues with the least real parts
resp.\ absolute values. Of course the non-zero eigenvalues occur in
complex conjugate pairs, hence we only consider one sign for the
imaginary part.

\TABLE[t]{\begin{tabular}{|c|c|c|c|c|c|}
\hline
lattice & & physical &
\multicolumn{3}{|c|}{number of configurations} \\
size & $\beta$ & volume & $\nu = 0$ & $\vert \nu \vert = 1$ 
& $\vert \nu  \vert = 2$  \\
\hline
\hline
$12^{4}$ & 6 & $(1.12 \fm)^{4}$ & 44 & 70 & 24  \\
\hline 
$10^{4}$ & 5.85 & $(1.23 \fm)^{4}$ & 74 & 112 & 84 \\
\hline
$8^{4}$ & 5.85 & $(0.98 \fm)^{4}$ & 80 & 63 & 28 \\
\hline
\end{tabular}%
\caption{The statistics of our simulations on three
lattice sizes.\label{stati}\label{tab1}}}

First we measured the leading non-zero eigenvalue on a $12^4$ lattice
at $\beta =6$.  Here the volume amounts to $(1.12 \fm)^{4}$.
The result can be represented by a histogram, which could then be
compared to figure~\ref{RMT}.  However, such a picture depends on the
(arbitrary) choice of the bin size in the histogram. This can be
avoided by plotting the ``cumulative density'', see e.g.\
ref.~\cite{NR}, which sums up all the entries up to the considered
value of $z$ --- it is normalized so that the full number of entries
corresponds to the cumulative density 1.  This way to plot the data
can be compared to the cumulative density according to the RMT
prediction,
\begin{equation}
\rho_{1,c}^{(\nu )}(z) = \frac{
\int_{0}^{z} \rho_{1}^{(\nu )}(z') dz'}{
\int_{0}^{\infty} \rho_{1}^{(\nu )}(z') dz'} \,.
\end{equation}
This comparison is done in figure~\ref{largeV} (top).  We recognize a
satisfactory agreement, especially for $\vert \nu \vert =1$, where we
have the largest statistics, see table~\ref{stati}.  In particular, we
do see the effect that the peak of the density --- resp.\ the interval
of steepest ascent of the cumulative density --- moves to larger
values of $z$ for increasing topological charge.  In this plot the
chiral condensate $\Sigma $ enters as the one free parameter, which
sets the energy scale. Our plot is shown for the value
\begin{equation}
\Sigma = (256 \MeV)^{3} \,,
\end{equation}
which provides optimal agreement with the RMT curves.

\FIGURE[t]{\epsfig{file=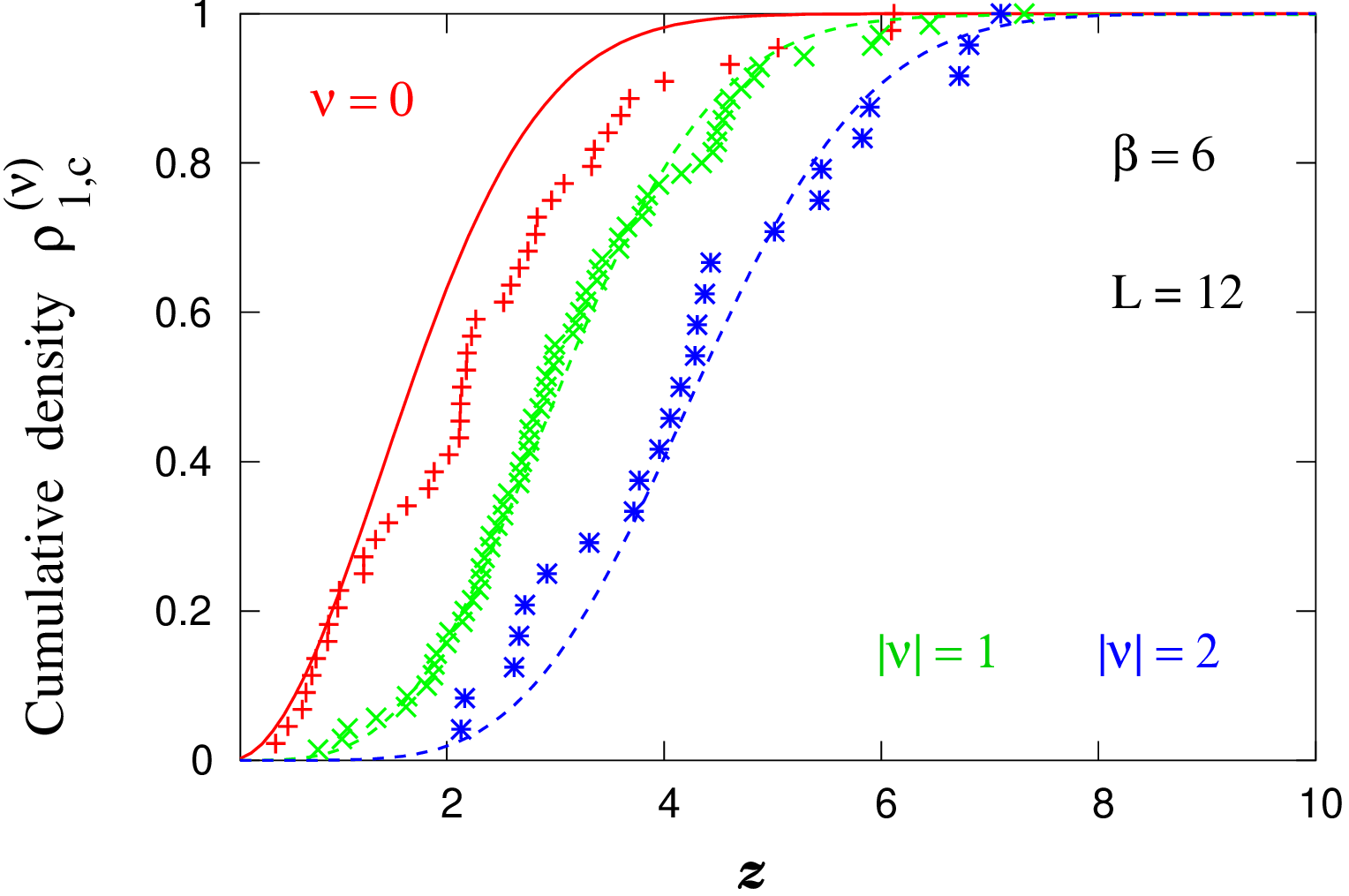,angle=270,width=.6\textwidth,clip=}
\epsfig{file=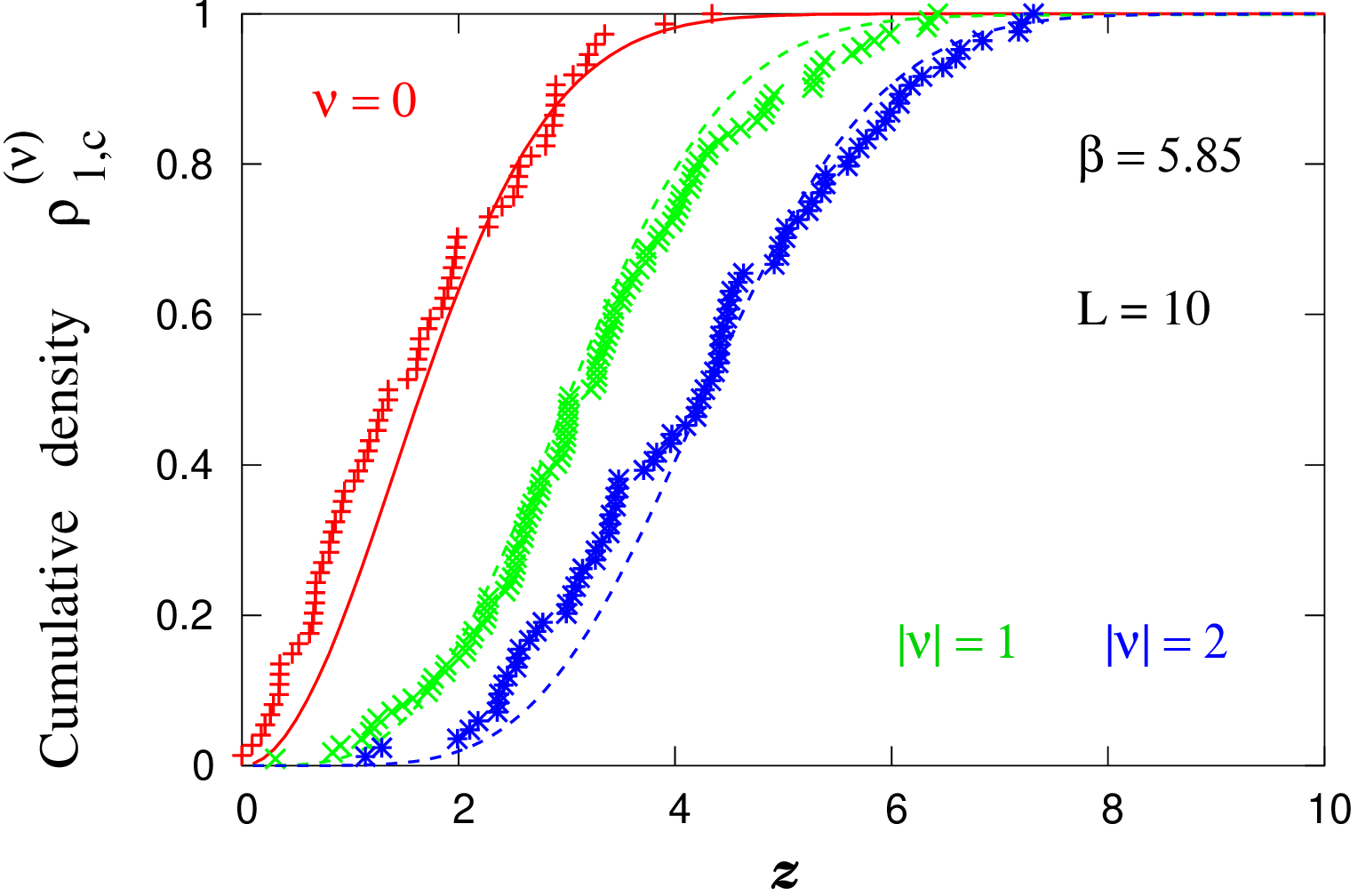,angle=270,width=.6\textwidth,clip=}%
\caption{The cumulative distribution of the lowest non-zero eigenvalue
at $\beta =6$ on a $12^{4}$ lattice (top), and on a $10^{4}$ lattice
at $\beta = 5.85$ (bottom).  We show the RMT predictions (lines) and
result for the indices $\vert \nu \vert =0,1,2$. In both cases, in
particular at $L=10$, we find a satisfactory agreement with RMT, if
the chiral condensate is chosen optimally. This requires $\Sigma =
(256 {\rm MeV})^{3}$ (top) resp.\ $\Sigma = (253 {\rm MeV})^{3}$
(bottom).  The $10^{4}$ lattice corresponds to a somewhat larger
physical volume.\label{largeV}\label{fig2}}}

Next we studied a $10^4$ lattice at $\beta =5.85$, which corresponds
to a somewhat larger physical volume of $V = (1.23 \fm)^{4}$.  Again
the results for the leading non-zero eigenvalue are shown for the
topological sectors $\vert \nu \vert =0, 1$ and $2$ in
figure~\ref{largeV} (bottom). Also here we find a reasonable agreement
with the RMT predictions, and the optimal value of the chiral
condensate is modified to $\Sigma = (253 ~ {\rm MeV})^{3}$.  RMT
predicts a (logarithmically) increasing $\Sigma (V)$ due to
quenching~\cite{loga}.  A conclusive verification of this behavior
(beyond possible lattice artifacts) would require further simulations.
In figure~\ref{FigEV2} we show our results for the cumulative density
of the next-to-leading eigenvalue at $\vert \nu \vert =0$, $1$ and
$2$, for both lattices discussed before.  Here we do not find a
convincing agreement with the RMT prediction in the first
ref. of~\cite{lowEV2}; note that the relevant values of $z$ are larger
compared to figure~\ref{fig2}.  Also the second non-zero eigenvalue
moves to larger values of $z$ if $\vert \nu \vert$ increases.

\FIGURE[t]{\epsfig{file=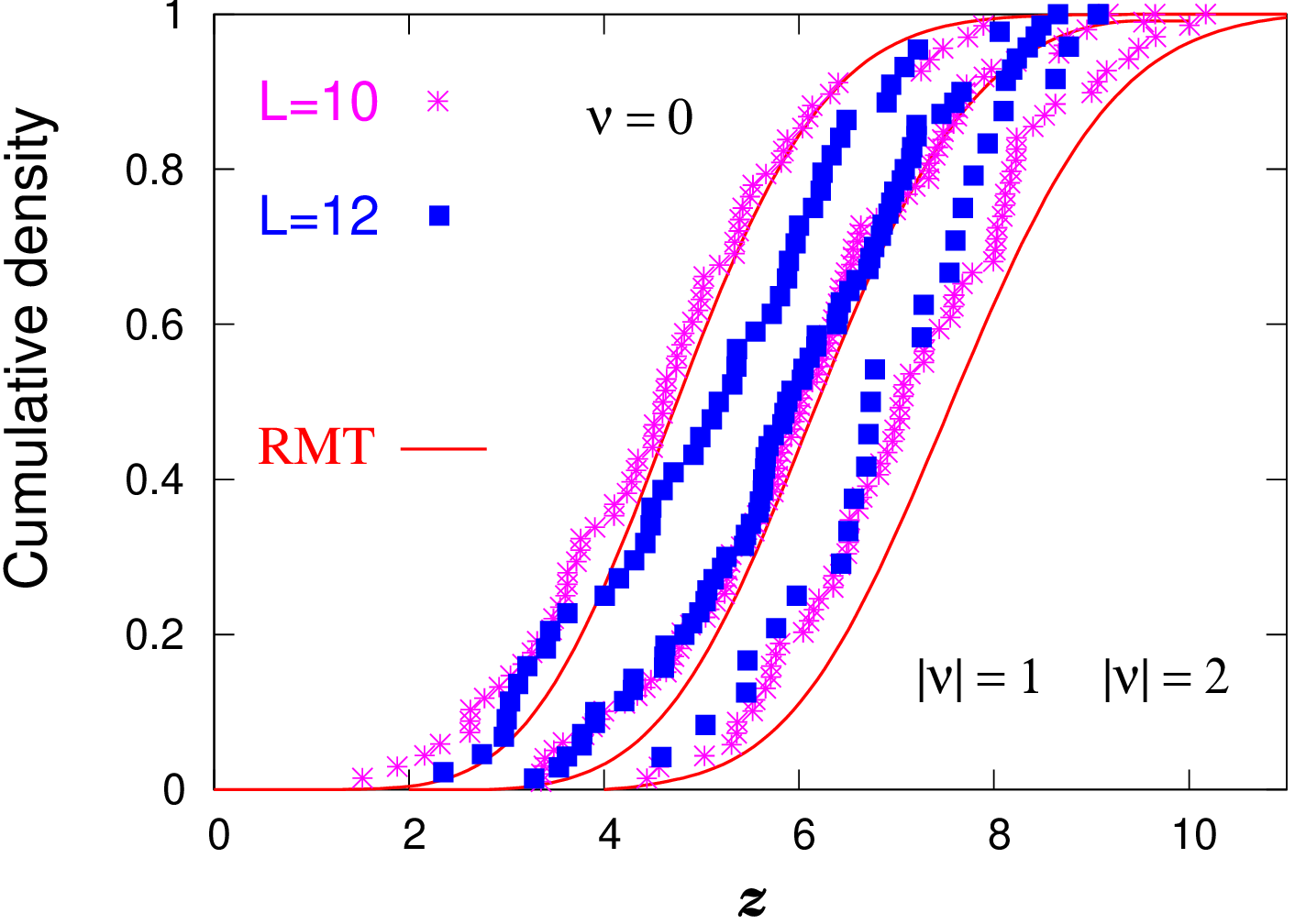,angle=270,width=.6\textwidth,clip=}%
\caption{The cumulative density of the second non-zero eigenvalue at
$L=12$, $\beta=6$ and at $L=10$, $\beta = 5.85$.  At $\vert \nu \vert
= 0,$ $1$ and $2$ we do not find a convincing agreement with the RMT
prediction (solid lines). Also the second eigenvalue moves to larger
values for increasing $\vert \nu \vert$.\label{FigEV2}\label{fig3}}}

Now we consider the \emph{eigenvalue density} without focusing
particularly on the first or on the second non-zero eigenvalue (though
we still omit the zeros).  Then we obtain for instance at $\vert \nu
\vert =1$ the densities shown in figure~\ref{microdense}. RMT predicts
an oscillating behavior (see first ref.\ in~\cite{lowEV}), which is
also plotted for comparison. We find a good agreement roughly up to
the second peak (as expected from figures~\ref{largeV}
and~\ref{FigEV2}). Then we are leaving the microscopic regime and
turn to the bulk; for the latter the density is supposed to rise as
$\lambda^{3}$ resp.\ $z^{3}$. This is nicely confirmed by our data; an
example is illustrated in figure~\ref{bulk}.

\FIGURE[t]{\epsfig{file=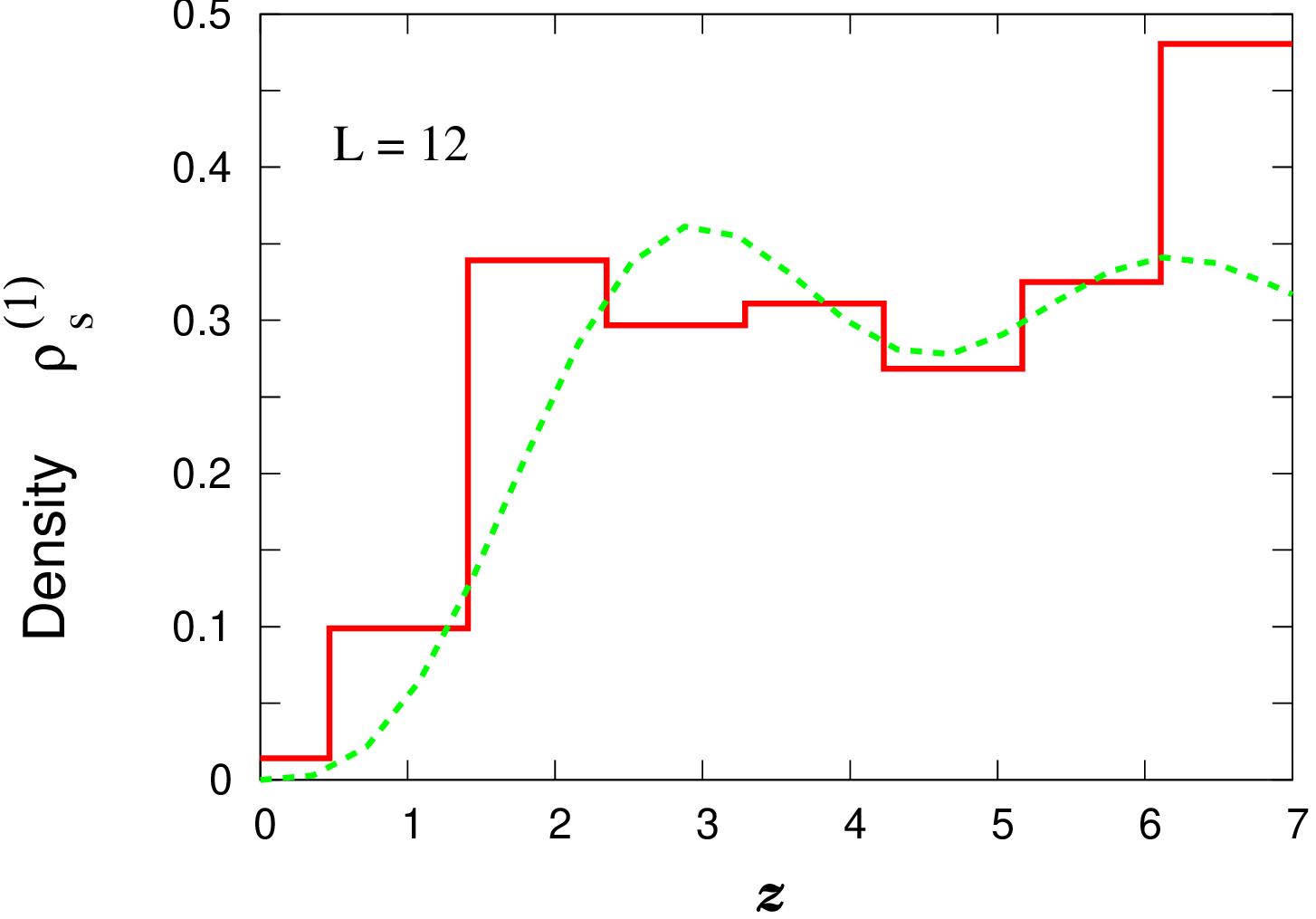,angle=270,width=.6\textwidth,clip=}
\epsfig{file=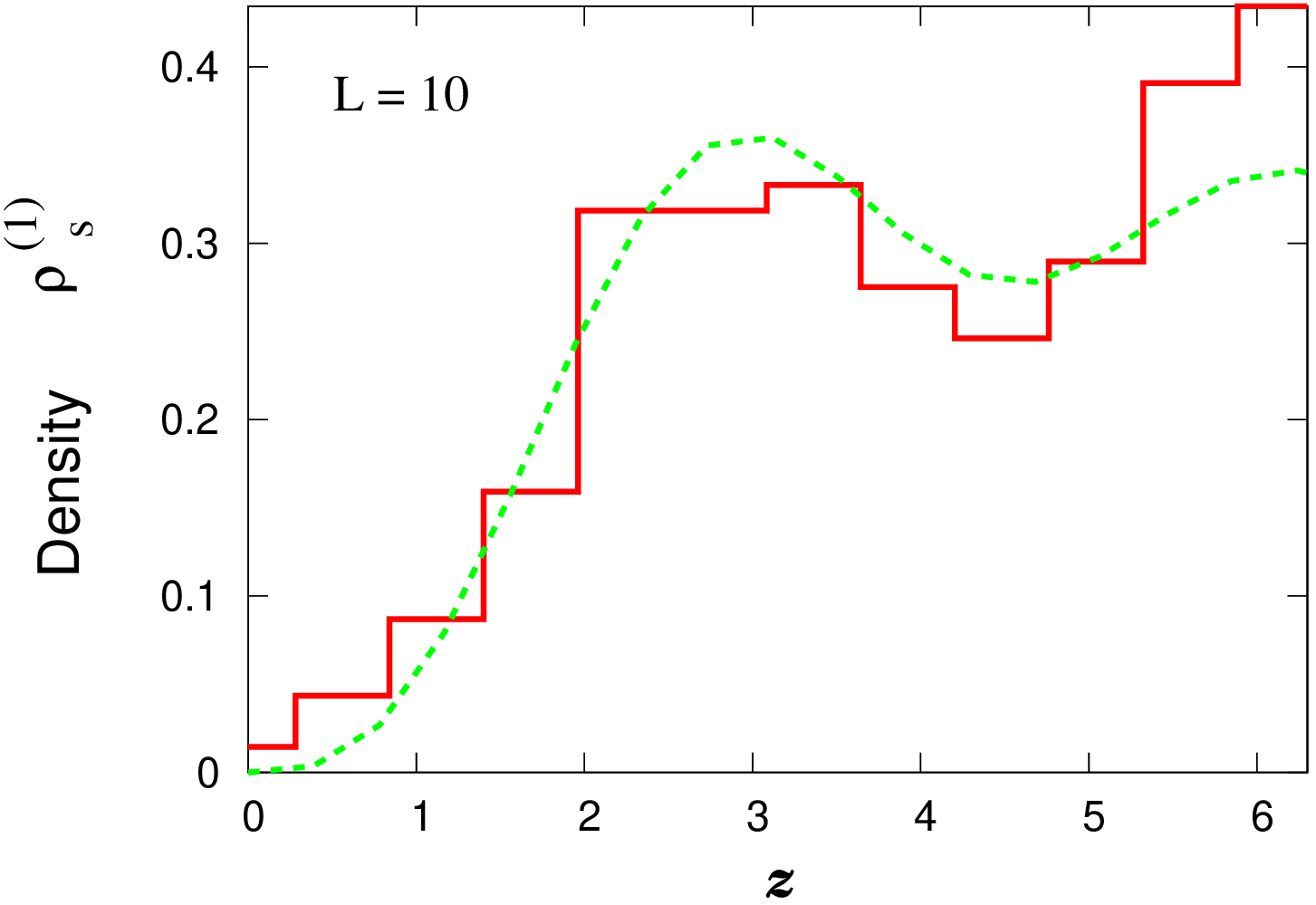,angle=270,width=.6\textwidth,clip=}
\caption{The density of eigenvalues of the Dirac operator in the
microscopic regime, for the $12^4$ lattice at $\beta =6$ (top) and for
the $10^4$ lattice at $\beta =5.85$ (bottom).  In both cases, we show
as an example the data for $\vert \nu \vert =1$, which follow the RMT
predicted oscillating behavior (dashed line) roughly up to the second
peak. If the oscillation is averaged to a plateau, its hight agrees
with the eigenvalue density at zero according to the Banks-Casher
relation, $\rho (z=0) = \rho (\lambda = 0)/\Sigma V = 1/
\pi$.\label{microdense}\label{fig4}}}

\FIGURE[t]{\epsfig{file=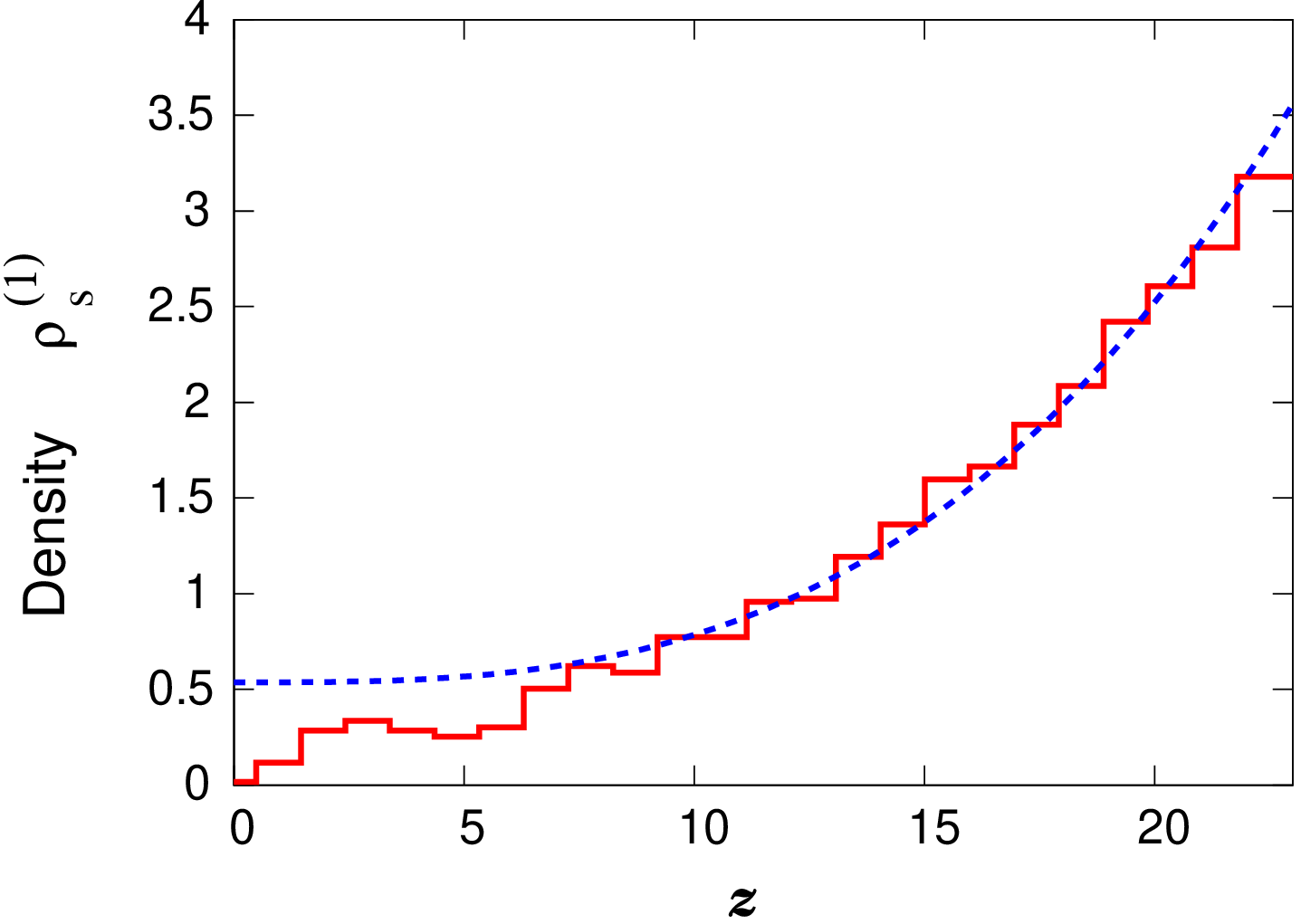,angle=270,width=.6\textwidth,clip=}%
\caption{The eigenvalue density on the $12^4$ lattice at $\beta =6$,
$\vert \nu \vert =1$. We see the transition from the microscopic
regime to the bulk, where the density follows the predicted increase
$c_{0} + c_{1}\lambda^{3}$ (dashed line).\label{bulk}\label{fig5}}}

Finally we consider a smaller volume: we use a $8^4$ lattice at $\beta
= 5.85$, which corresponds to $V = (0.98 \fm)^{4}$.
Figure~\ref{L8b5} shows that in this case there is a clear
disagreement with the RMT prediction; in particular the sensitivity of
the peak to $\vert \nu \vert $ is lost to a large extent (beyond very
small values of $z$).  This is in qualitative agreement with the
results of ref.~\cite{Bern} on the $4^4$ lattice.  The observation
that the volume has to be $V \gsi (1.2\fm)^{4}$ can be viewed as
an empirical determination of the Thouless energy.
\FIGURE[t]{\epsfig{file=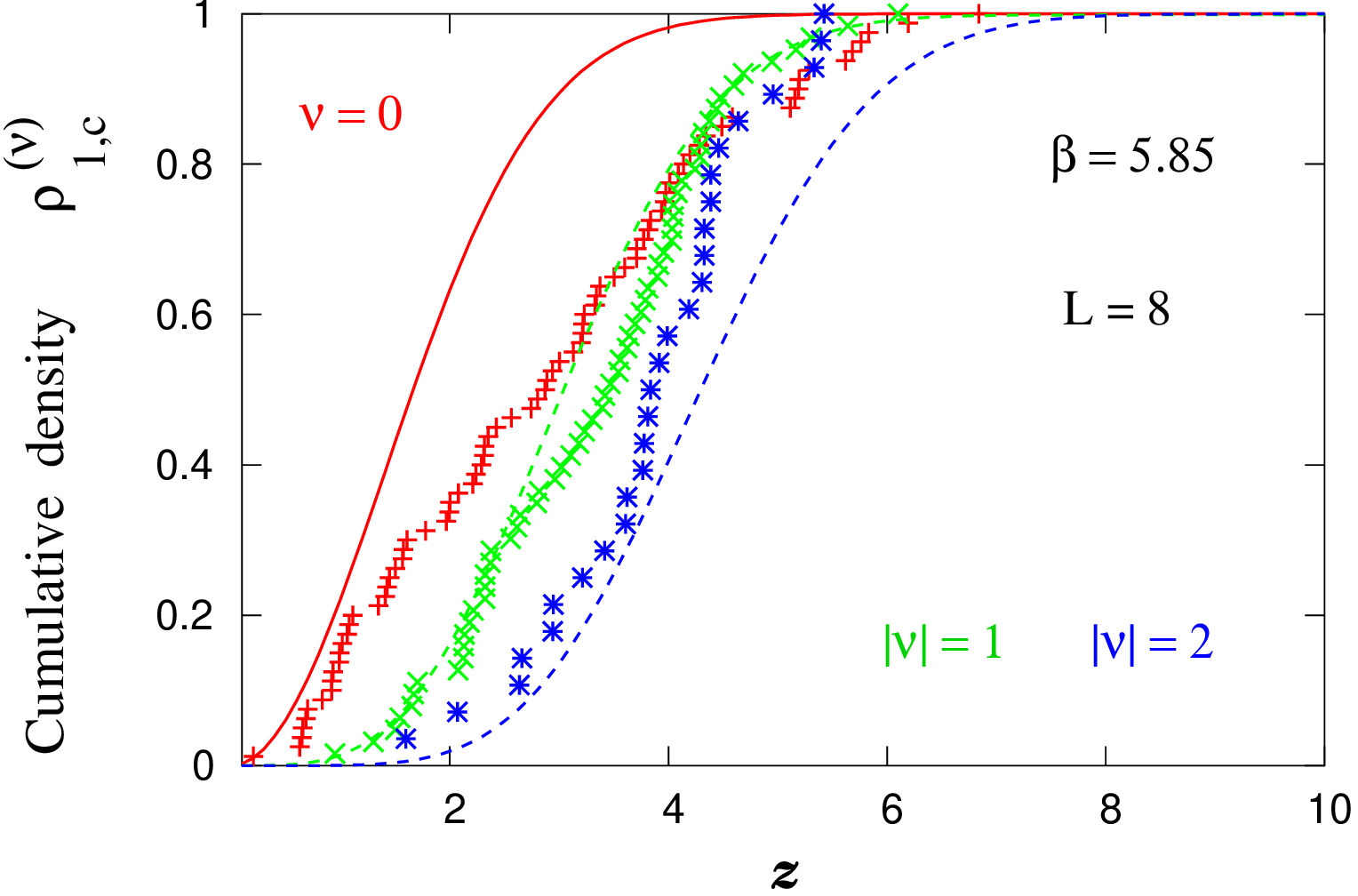,angle=270,width=.6\textwidth,clip=}%
\caption{The cumulative density of the lowest non-zero eigenvalue at
$\beta = 5.85$ on a $8^{4}$ lattice.  The data from different
topological sectors are close to each other (beyond very small $z$
values), hence in this small volume they do not agree with the RMT
curves.\label{L8b5}\label{fig6}}}

At this point, we would like to add that the above results tend to
look similar for the leading eigenvalues of the hermitean overlap
Dirac operator
\begin{equation}
H_{\rm ov} = \gamma_{5} D_{\rm ov} \,. 
\end{equation}
(Of course, in this case the 
stereographic mapping is not needed.) As an example we show the
cumulative density of the first positive (non-zero) eigenvalue of
$H_{\rm ov}$ at $L=10$, $\beta = 5.85$ in figure~\ref{hermi}.
\FIGURE[t]{\epsfig{file=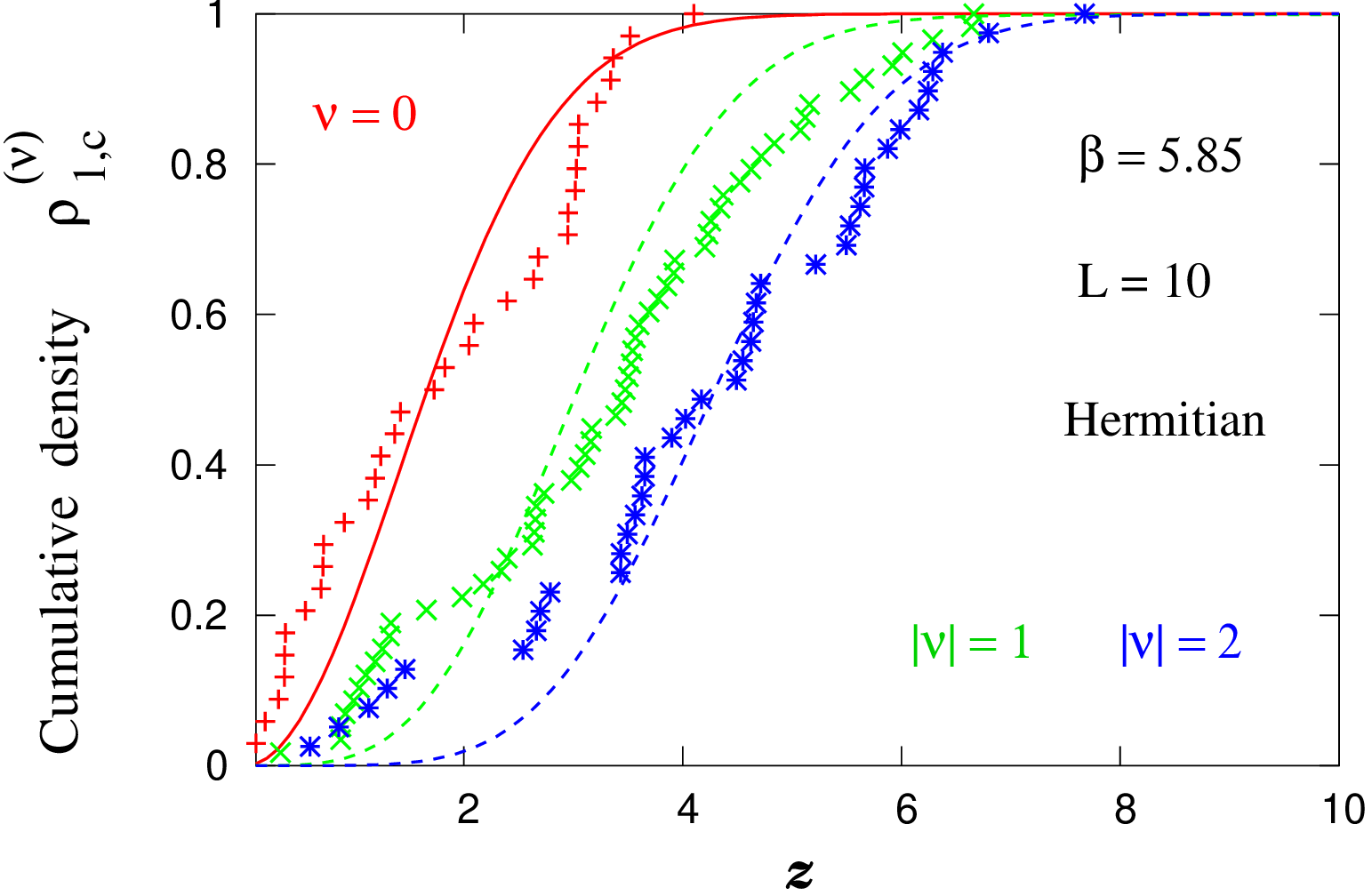,angle=270,width=.6\textwidth,clip=}%
\caption{The cumulative density of the lowest non-zero eigenvalue of
the hermitean overlap Dirac operator $H_{\rm ov} = \gamma_{5} D_{\rm
ov}$ on the $10^{4}$ lattice at $\beta
=5.85$.\label{hermi}\label{fig7}}}

\TABLE[t]{\begin{tabular}{|c|c|c|c|c|c|c|}
\hline
lattice &      & 
\multicolumn{3}{|c|}{confidence level}
& optimal & \\
size & $\beta$ & $\nu = 0$ & $\vert \nu \vert = 1$ 
& $\vert \nu \vert = 2$ & $\Sigma $ & $\chi_{t} \, r_{0}^{4}$ \\
\hline
\hline
$12^{4}$ & 6 & 0.003 & 0.73 & 0.79 
& $(256~{\rm MeV})^{3}$ & 0.063 \\
\hline 
$10^{4}$ & 5.85 & 0.03 & 0.48 & 0.10 
& $(253~{\rm MeV})^{3}$ & 0.078 \\
\hline
\end{tabular}%
\caption{The results on the two lattices which turned out to be
sufficiently large for the data to be compared with the RMT
predictions, for the suitable values of $\Sigma$ used in
figure~\ref{largeV}. We also display the statistical confidence level
according to the Kolmogorov-Smirnov test, and our results for the
topological susceptibility $\chi_{t}$ ($r_{0}$ is the Sommer
parameter).\label{resu}\label{tab2}}}

For completeness we report our statistics in table~\ref{stati} and
summarize the results in table~\ref{resu}.  In that table we also
include an estimate for the topological susceptibility, where some
configurations with higher charges contribute as well. We present the
dimensionless quantity $\chi_{t} r_{0}^{4}$, where $\chi_{t} = \langle
\nu^{2} \rangle / V$ is the susceptibility and $r_{0}$ is the Sommer
parameter obtained from the static quark-antiquark
potential~\cite{Som}.  The results are well consistent with the values
in the literature (for an overview, see ref.~\cite[figure~18]{Bern}),
although our statistics is clearly too small for that quantity.
(Hence we do not try to estimate errors on $\Sigma$ and
$\chi_{t}r_{0}^{4}$.)  Finally table~\ref{resu} also gives the results
of the Kolmogorov-Smirnov test~\cite{NR} for the confidence level of
the results obtained, if one assumes the RMT probability distribution
to be correct. The corresponding number described the probability for
the observed deviation from the theory to occur with the given
statistics. This test is just designed for the cumulative density.
The results suggest that the volume $(1.12 \fm)^{4}$ is still somewhat
small, so that the topologies are still not fully separated, but we
see that this process sets in at least for the first non-zero
eigenvalue on volumes larger than that.

\subsection{Unfolded distribution}

We now turn to another way of comparing our lattice data to a
conjecture from RMT. This new evaluation allows us to take all our
non-zero eigenvalues into account (for one sign of the imaginary part,
i.e.\ up to 50 for each configuration).

We build from all the eigenvalues of all configurations the ``unfolded
distribution'' as described for instance in
ref.~\cite{SCRI}\footnote{A different, more systematic notion of
unfolding is described in ref.~\cite{review}.}  (for generalities, see
ref.~\cite{RMTbooks}).  To this end, we first numerate all available
non-zero eigenvalues with positive imaginary part in each
configuration in ascending order, given by the angle in the
Ginsparg-Wilson circle. Then we put all these eigenvalues from all
configurations together and numerate them again in ascending order.
Now we consider pairs of eigenvalues from the same configuration,
which follow immediately one after the other in the original
numeration. If they differ in the global numeration by $k$, then $k /
N_{{\rm conf}}$ is the ``unfolded level spacing'', where $N_{{\rm
conf}}$ is the number of configurations involved.

In figure~\ref{unfold} we show our results for
$L=12$, at $\beta =6$ and for $L=8$, at $\beta =5.85$.
The histograms are compared to RMT conjectures for different
groups and representations, and we can confirm a good agreement with
the Wigner distribution predicted for $\SU(3)$ in the fundamental 
representation. Considering also ref.~\cite{SCRI} we conclude that this
property holds over a large range of volumes. Since we include
all eigenvalues here, our statistics is much larger than in the
plots for the microscopic regime. Note, however,
that this analysis is not sensitive to the topology any more.
\FIGURE[t]{\epsfig{file=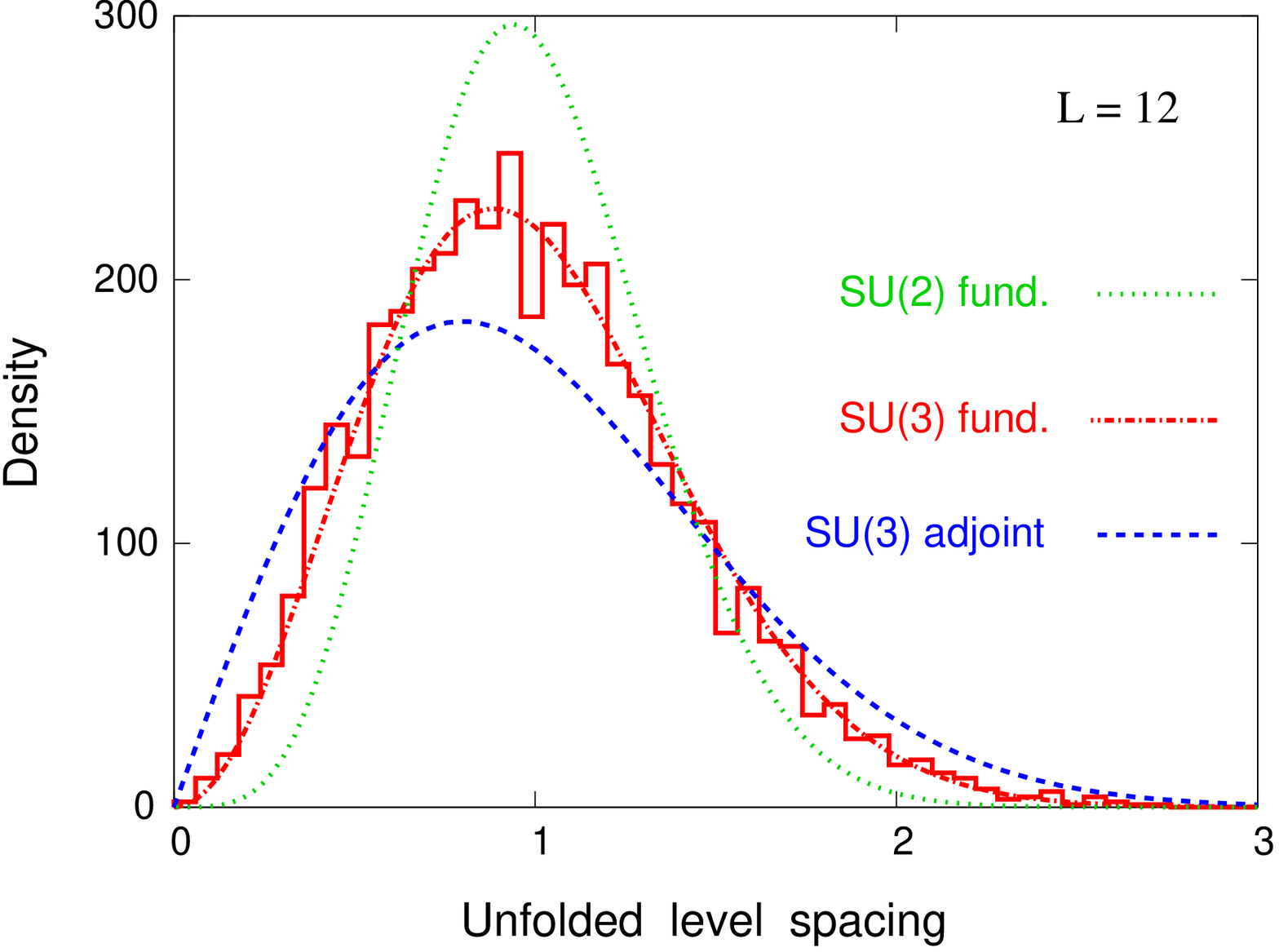,angle=270,width=.6\textwidth,clip=}
\epsfig{file=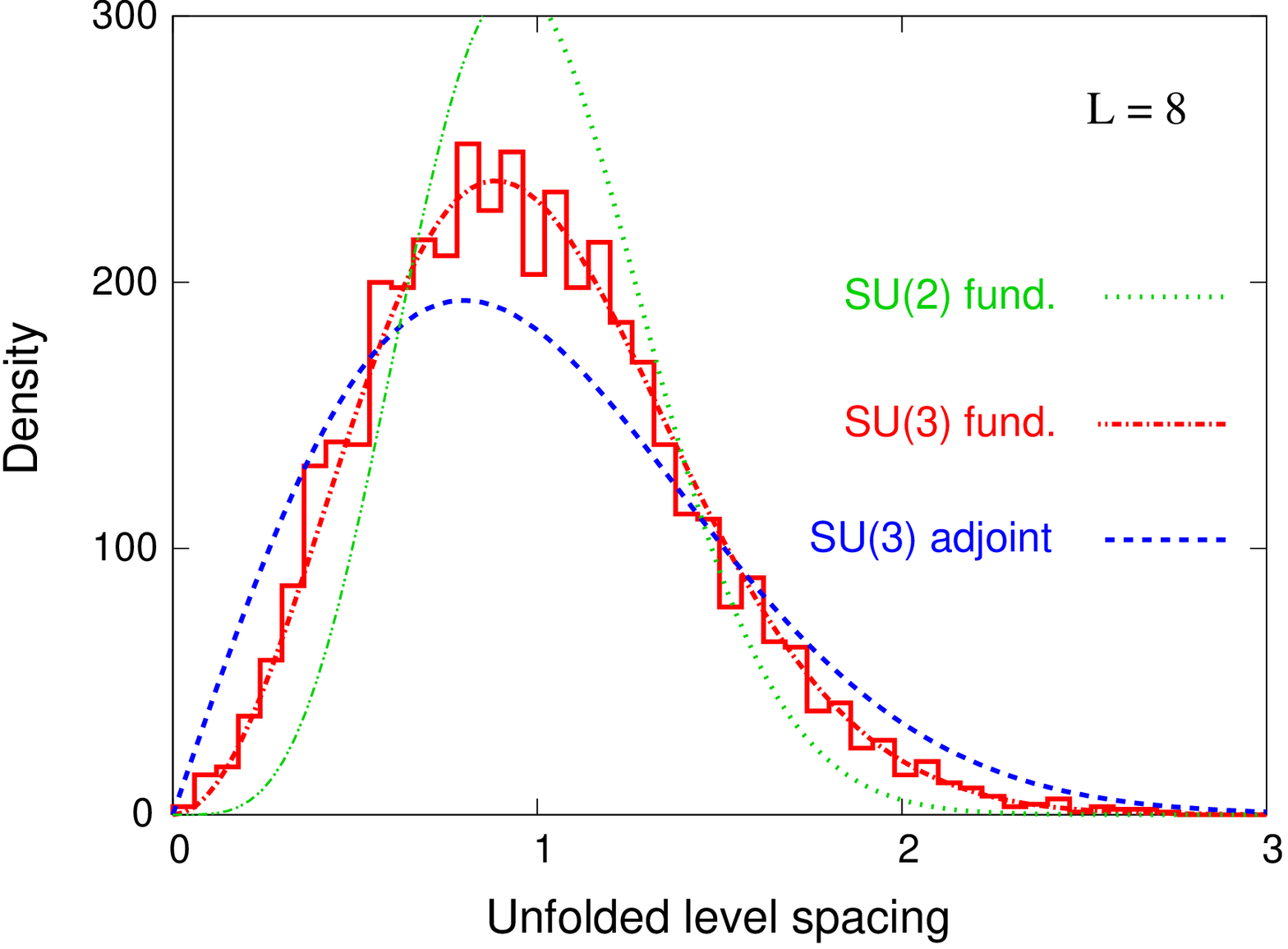,angle=270,width=.6\textwidth,clip=}
\caption{The unfolded level spacing distribution for $\beta =6$ on a
$12^{4}$ lattice (top), and for $\beta =5.85$ on a $8^{4}$ lattice
(bottom).  We find a convincing agreement with the RMT conjecture for
the $\SU(3)$ group in the fundamental representation that we are
using. The same also holds for the $10^4$ lattice at $\beta = 5.85$.
(The curves are normalized according to the number of entries in the
statistics.)\label{unfold}\label{fig8}}}

Moreover the unfolded distribution only tests the correct symmetry
group that is to be taken in RMT. There is no physics information such
as values for the condensate or the pion decay constant, because one
does not keep track of a physical scale.  Nevertheless, the unfolded
distribution provides a non-trivial test for RMT.

\section{Conclusions}

We performed a lattice study of quenched QCD using the overlap
operator to test RMT predictions for the eigenvalue distributions of
the Dirac operator.  In this pilot study we worked at $\beta$ values
large enough so that the overlap formula can safely be applied as a
solution of the GWR. By using lattices of size $8^{4}$, $10^{4}$ and
$12^{4}$ we were able to reach physical volumes which are large enough
to test RMT.

In the microscopic regime of the very small eigenvalues we observe
agreement with the predictions by RMT applied to the $\eps$-regime of
chiral perturbation theory, \emph{if} the physical volume is large
enough; $V \gsi (1.2 \fm)^{4}$ seems to be sufficient to capture at
least the leading non-zero eigenvalue.  On smaller physical volumes
this agreement --- and in particular the sensitivity to the topology
--- disappears.

If we consider the unfolded level spacing distribution, however, we
obtain a good agreement with the RMT conjecture, even down to a small
physical volume.

The eigenvalue distributions in figure~\ref{largeV} show interesting
characteristic features.  In topological charge sector zero the
probability for a very low non-zero eigenvalue is
non-negligible. Therefore such very low-lying modes will appear in
simulations.  This might have serious consequences for numerical
studies when the quark mass is lowered too much.  In particular, this
behavior might render simulations in the $\epsilon$-regime
problematic. Indications of these difficulties were presented in
ref.~\cite{condensate}, where also an analysis of the influence of the
lowest mode in the $\nu = 0$ sector on the measurement of the chiral
condensate was given.  It was reported that the condensate is very
hard to measure in topological charge sector zero, in agreement with
the results presented here.

On the other hand, the situation is clearly better in a non-trivial
topology. Eigenvalues close to zero are strongly suppressed and a much
better signal can be expected.

Note that this phenomenon is not restricted to the quenched
approximation considered here but holds also in the full theory.
Thus, it can be expected that simulations close to physical values of
the pion mass may become extremely expensive. Clearly, contact to
chiral perturbation theory is then mandatory to extrapolate to
physical pion masses.

\subsection{Outlook}

We still want to proceed to larger physical volumes. This will
hopefully lead to further precision in the agreement for the leading
non-zero eigenvalues in the different topological sectors.  The larger
volume might provide a histogram for the spectral density, which
follows the microscopic RMT for several oscillations. With an
increased statistics --- which may be accessible using the optimized
algorithmic techniques described in ref.~\cite{algo} --- we could
evaluate the spectral correlators as well.  Further values for $\Sigma
(V)$ would also allow us to extract an estimation for
$F_{\pi}$~\cite{loga}.  Finally we are also about to extend this
analysis to a hypercube overlap fermion, which is described in
ref.~\cite{HF2}.  This is a different Ginsparg-Wilson fermion, and its
study will provide a more complete picture of the spectral behavior of
this class of lattice Dirac operators.

\acknowledgments

We would like to thank Poul Damgaard for his help with the
interpretation of the formulae in ref.~\cite{lowEV2}.  We are also
indebted to him as well as Gernot Akemann, Michael M\"{u}ller-Preu\ss
ker, Jacques Verbaarschot and Tilo Wettig for useful discussions and
many helpful comments.  K.J.\ wants to thank Leonardo Giusti, Pilar
Hern\'andez, Martin L\"uscher, Peter Weisz and Hartmut Wittig for
helpful discussions about RMT and the interpretation of the eigenvalue
distributions.  We further thank Kei-ichi Nagai for communication
regarding the numerical work. The computations have been performed at
the Konrad Zuse Zentrum in Berlin and at the Forschungszentrum
J\"{u}lich.

This work was supported in part by the DFG Sonderforschungsbereich
Transregio 9, ``Computergest\"{u}tzte Theoretische Teilchenphysik''.

\end{document}